\documentclass[twocolumn, trackchanges, resetfootnote]{aastex701}

\usepackage{amsmath,amsfonts,amssymb}
\usepackage{bm}
\usepackage{graphics}
\usepackage{graphicx}
\usepackage{here} 
\usepackage{type1cm}
\usepackage{multirow}
\usepackage{float}
\usepackage{natbib}

\hypersetup{
unicode=true,
setpagesize=false,
bookmarksnumbered=true,
bookmarksopen=true,
colorlinks=true,
linkcolor=blue,
citecolor=blue,
urlcolor=blue
}

\def\Cone{\hbox{C\,{\scriptsize I}}}

\shorttitle{New Precise Measurement of the Cosmic Microwave Background Radiation Temperature}
\shortauthors{Kotani et al.}
\graphicspath{{./}{figures/}}
\received{May 30, 2025}
\revised{September 18, 2025}
\accepted{September 19, 2025}
\submitjournal{ApJ}
\begin{document}

\title{A New Precise Measurement of the Cosmic Microwave Background Radiation Temperature at $z=0.89$ Toward PKS1830--211}

\correspondingauthor{Tatsuya Kotani}
\email{sci.tatsu.729@keio.jp}

\author[orcid=0009-0006-9842-4830,gname=Tatsuya,sname=Kotani]{Tatsuya Kotani}
\affiliation{School of Fundamental Science and Technology, Graduate School of Science and Technology, Keio University, 3-14-1 Hiyoshi, Kohoku-ku, Yokohama, Kanagawa 223-8522, Japan}
\email{sci.tatsu.729@keio.jp}

\author[orcid=0000-0002-5566-0634,gname=Tomoharu,sname=Oka]{Tomoharu Oka} 
\affiliation{School of Fundamental Science and Technology, Graduate School of Science and Technology, Keio University, 3-14-1 Hiyoshi, Kohoku-ku, Yokohama, Kanagawa 223-8522, Japan}
\affiliation{Department of Physics, Institute of Science and Technology, Keio University, 3-14-1 Hiyoshi, Kohoku-ku, Yokohama, Kanagawa 223-8522, Japan}
\email{tomo@phys.keio.ac.jp}

\author[orcid=0000-0003-2735-3239,gname=Rei,sname=Enokiya]{Rei Enokiya} 
\affiliation{National Astronomical Observatory of Japan, 2-21-1 Osawa, Mitaka, Tokyo 181-8588, Japan}
\email{rei.enokiya@nao.ac.jp}

\author[gname=Kazuki,sname=Yanagihara]{Kazuki Yanagihara} 
\affiliation{School of Fundamental Science and Technology, Graduate School of Science and Technology, Keio University, 3-14-1 Hiyoshi, Kohoku-ku, Yokohama, Kanagawa 223-8522, Japan}
\email{lakalakalove.uver@keio.jp}

\author[orcid=0000-0003-4732-8196,gname=Miyuki,sname=Kaneko]{Miyuki Kaneko} 
\affiliation{School of Fundamental Science and Technology, Graduate School of Science and Technology, Keio University, 3-14-1 Hiyoshi, Kohoku-ku, Yokohama, Kanagawa 223-8522, Japan}
\email{miyukikaneko@keio.jp}

\author[gname=Ryo,sname=Ariyama]{Ryo Ariyama} 
\affiliation{School of Fundamental Science and Technology, Graduate School of Science and Technology, Keio University, 3-14-1 Hiyoshi, Kohoku-ku, Yokohama, Kanagawa 223-8522, Japan}
\email{sirius@keio.jp}

\begin{abstract}
  In this study, we analyzed millimeter-wave data toward the quasar PKS1830--211 observed with the Atacama Large Millimeter/submillimeter Array to obtain absorption spectra of HCN {\it J}=2--1, {\it J}=3--2, {\it J}=4--3, and {\it J}=5--4 lines at the cosmological redshift of $z\!=\!0.89$. We confirmed multiple absorption components in each spectrum, and the two lower-{\it J} lines are highly saturated at velocity centers of the two most prominent components. 
  The effect of uncertainty in the continuum covering factor was carefully assessed using a Monte Carlo approach.
  We also accounted for systematic uncertainties in the HCN {\it J}=4--3 and {\it J}=5--4 absorption due to time variability during an intense flaring event of the quasar. Assuming local thermodynamic equilibrium and neglecting molecular collisions, we derived the excitation temperature profiles and their uncertainties in the optically thin regime.
  We determined the cosmic microwave background radiation temperature at $z = 0.89$ to be ${5.13\pm0.06\,\mathrm{K}}$ by taking a weighted average of calculated excitation temperatures; this is highly consistent with that expected from the standard model of the universe.    
\end{abstract}

%% You will be asked to selected these concepts during the submission process but this old "keyword" functionality is maintained in case authors want to include these concepts in their preprints.

\section{Introduction} \label{sec:intro}
The cosmic microwave background (CMB; \citealp{Penzias_1965, Dicke_1965}) is an almost uniform and isotropic radiation field that has a perfect black body spectrum at $\sim2.7\,\mathrm{K}$.  This is among the strongest observational evidence for the Big Bang theory, which is interpreted as a snapshot of the hot, dense, ionized universe before the recombination epoch at 380,000 years after the Big Bang.  The CMB temperature decreases with cosmic time as the universe expands adiabatically; this means that distant galaxies exhibit themselves immersed in earlier (warmer) CMB radiation.  The standard Big Bang model predicts that the CMB temperature ($T_\mathrm{CMB}$) evolves with the cosmological redshift ($z$) of the observed distant object as 
\begin{equation}\label{eq:tcmb}
T_\mathrm{CMB}(z)=T_0\,(1+z),
\end{equation}
where $T_0$ ($=2.72548\pm0.00057\,\mathrm{K}$; \citealp{Fixsen_2009}) represents the current CMB temperature.  

Equation~(\ref{eq:tcmb}) can be tested by measuring $T_\mathrm{CMB}(z)$ at various redshifts.  Thus far, a number of $T_\mathrm{CMB}(z)$ measurements have been performed using multifrequency thermal Sunyaev--Zel'dovich observations ($z\!<\!1$; \citealp{Battistelli_2002, Horellou_2005, Luzzi_2009, Hurier_2014, Saro_2014, deMartino_2015}) and the excitation analyses of atomic or molecular absorption lines toward quasars ($z\!>\!1$; \citealp[][]{Ge_1997, Srianand_2000, Molaro_2002, Cui_2005, Klimenko_2020}).  The measurement results are roughly consistent with the standard model; however, the measurement uncertainties exceed several percent at $z\!\gtrsim\!1$.  
 
PKS1830--211 is a quasar at $z\!=\!2.507$ \citep{Lidman_1999}, which shows a rich absorption line system at $z\!=\!0.88582$ \citep{Wiklind_1996}.  This absorption line system is attributed to an almost face-on spiral galaxy in the foreground \citep{Winn_2002, Courbin_2002}.  PKS1830--211 is observed as two point-like sources (NE and SW) at millimeter wavelengths because of the gravitational lensing effect caused by the foreground spiral galaxy.  More than 60 molecular species and 15 rare isotopologues have been detected in the SW component (\citealp[][]{Wiklind_1996, Muller_2006, Muller_2011, Muller_2014, Tercero_2020, Muller_2024}).  \citet{Muller_2013} determined
$T_\mathrm{CMB}(z\!=\!0.89)\!=\!5.08\pm 0.10 \,\mathrm{K}$ (errors are at 68\% confidence level) using the 10 different molecular absorption line spectra in the millimeter wavelength toward the PKS1830--211 SW obtained with the Australia Telescope Compact Array.  This is currently the best estimate of $T_\mathrm{CMB}$ at this redshift, and it is consistent with $5.14\,\mathrm{K}$, which is expected from the standard Big Bang model within the uncertainty.  

In this study, we report on the new estimation of $T_\mathrm{CMB}(z\!=\!0.89)$ from the millimeter-wave data toward PKS1830--211 SW observed with the Atacama Large Millimeter/submillimeter Array (ALMA).  Only four HCN spectra with high signal-to-noise ratio were used in the analysis to minimize uncertainty, and the effect of continuum covering factor, time variability, and column density inhomogeneity were carefully evaluated (Sect.~\ref{sec:analy}).  Increasing the number of molecular species used in the analysis does not necessarily improve accuracy because it increases the number of uncertain physical parameters.  The deviation from the standard Big Bang model is more tightly constrained by the new, more precise value of $T_{\mathrm{CMB}}(z = 0.89)$, corresponding to an epoch when the universe was less than half of current age (Sect.~\ref{sec:resdis}).

\section{Data Processing} \label{sec:data}
We retrieved ALMA bands 3--6 data toward PKS1830--211 from the ALMA Science Archive\footnote{\url{https://almascience.eso.org/aq/}}.  Project codes of the datasets used are 2013.1.01099.S (PI: K. Kawaguchi), 2017.1.01119.S (PI: S. Wallstr\"om), 2018.1.00051.S (PI: S. Muller), and 2018.1.00692.S (PI: I. Mart\'i-Vidal).  They contain the {\it J}=2--1, {\it J}=3--2, {\it J}=4--3, and {\it J}=5--4 lines of hydrogen cyanide (HCN) at $z\!=\!0.89$, respectively.  

Data calibration and reduction were performed by the standard procedure within the CASA reduction package\footnote{\url{http://casa.nrao.edu/}}.  Quasars J1924--2914 or J2000--1748 were used for calibrating the bandpass and flux, and J1832--2039 was used for phase calibration.  Subsequently, we used the \texttt{mstransform} task in CASA to extract spectral windows with bandwidths of 1.9 GHz (0.94 GHz for band 4) that include the HCN absorption lines.

The spectra toward both NE and SW components of PKS1830--211 were extracted by the Python-based package $\texttt{UVMULTIFIT}$ \citep{MartiVidal_2014}.  This procedure includes two steps: peak position determination and spectrum construction.  The positions of these two components were determined by fitting two two-dimensional $\delta$-functions to all visibilities except for frequency channels where the absorption lines are visible.  Subsequently, we fitted two two-dimensional $\delta$-functions at fixed positions to the visibilities of each spectral channel for obtaining the spectra.  We use the SW spectrum for the following analyses because it contains considerably richer molecular lines than that of the NE spectrum.  

The frequency in the obtained spectrum was converted to the radial velocity using the relativistic Doppler relationship $V/c\!=\!\{1-({\nu}/{\nu_0})^2\}/\{1+({\nu}/{\nu_0})^2\}$, where $c$ and $\nu_0$ represent the light speed and rest frequency, respectively.  The barycentric velocity reference frame was adopted during the \texttt{mstransform} task, which regridded the spectral axis to match the minimum velocity resolution specified in the ALMA archive. The resulting velocity resolutions are 2.672, 1.179, 1.561, and 2.728 $\mathrm{km\,s^{-1}}$ for the HCN $J$=2--1, $J$=3--2, $J$=4--3, and $J$=5--4 transitions, respectively.
The redshift of the absorber ($z\!=\!0.88582$) corresponds to the systemic velocity of $V_\mathrm{sys}\!=\!1.6820\!\times\!10^{5}$ $\mathrm{km\,s^{-1}}$.  The velocities presented in this paper are offsets from this $V_\mathrm{sys}$.

\section{Analyses} \label{sec:analy}
\subsection{Fundamentals}\label{subsec:fundamentals} 
We assume one uniform cloud overlapping in the foreground on the quasar line of sight.  If we observe an absorption depth ($\Delta{I}$) against the continuum level ($I_\mathrm{c}$), the line optical depth can be calculated by 
\begin{equation}\label{eq:taudef}
  \tau_\nu=-\ln\left(1-\frac{1}{f_\mathrm{c}}\frac{\Delta{I}}{I_{\mathrm{c}}}\right),
\end{equation}
where $f_\mathrm{c}$ represents the continuum covering factor defined by $\Omega({\rm quasar}\cap{\rm cloud})/\Omega({\rm quasar})$.  

Here, we limit our discussion to the pure rotational lines of molecules.  The optical depth of the $J\!\rightarrow\!J+1$ absorption line is related to the column densities of the molecule at the lower and upper levels by 
\begin{eqnarray}
\int{\tau_{J+1,J}}\sqrt{\dfrac{1+{V}/{c}}{1-{V}/{c}}}\frac{1}{\left(1+{V}/{c}\right)^{2}}dV = \quad\quad\quad   \nonumber \\
=\frac{c^3A_{J+1,J} N_{J+1}}{8\pi\nu_{J+1,J}^3} \left[\frac{g_{J+1}N_{J}}{g_{J} N_{J+1}}-1\right], 
\end{eqnarray}
where $\nu_{J+1,J}$ represents the frequency of $J\!\leftrightarrow\!J+1$ transition, $A_{J+1,J}$ represents the Einstein A coefficient, $\tau_{J+1,J}$ represents the optical depth per unit velocity width, $g_{J}\!\equiv\!2J+1$ represents the statistical weight of the level $J$, $N_{J}$ represents the column density of the molecule at the level $J$, and $h$ and $k_\mathrm{B}$ represent the Planck and Boltzmann constants, respectively. 

Under the local thermodynamic equilibrium condition, ${N_{J}}$ (and ${N_{J+1}}$) can be rewritten by the excitation temperature ($T_{\mathrm{ex}}$) as
\begin{equation}\label{eq:tobesolved}
N_{J} = \frac{N}{ Z(T_{\mathrm{ex}})}  g_{J} \mathrm{exp} \left(-\frac{E_{J}}{ k_\mathrm{B}T_{\mathrm{ex}} }\right),
\end{equation}
where $N$ represents the total column density of the molecule, $Z(T_{\mathrm{ex}})\!\equiv\!\sum{g_{J}\mathrm{exp}(-E_{J}/{k_\mathrm{B}T_{\mathrm{ex}} }) }$ 
represents the partition function, and $E_{J}$ represents the energy of level $J$.  Using $E_{J+1}\!-\!E_{J}\!=\!h\nu_{J+1,J}$, we have
\begin{eqnarray}\label{eq:tobesolved}
  &&\frac{8\pi\nu_{J+1,J}^3}{c^3A_{J+1,J}g_{J+1}}\int{\tau_{J+1,J}}\sqrt{\dfrac{1+{V}/{c}}{1-{V}/{c}}}\frac{1}{\left(1+{V}/{c}\right)^{2}}dV \quad\quad \nonumber \\
  && =\frac{N}{Z(T_{\mathrm{ex}})}\mathrm{exp}\left(-\frac{E_{J+1}}{k_\mathrm{B}T_{\mathrm{ex}}}\right)\left[\exp\left({\frac{h\nu_{J+1,J}}{k_\mathrm{B}T_{\mathrm{ex}}}}\right)-1\right]. 
\end{eqnarray}
Equations (\ref{eq:taudef}) and (\ref{eq:tobesolved}) are the fundamentals of the analyses in this paper.

\subsection{$T_{\rm CMB}$ Calculations}\label{subsec:calculation} 
The excitation temperature $T_{\mathrm{ex}}$ and total column density of molecule $N$ can be calculated by fitting the right side of equation~(\ref{eq:tobesolved}) to the left side.  We need at least two transitions for determining these two unknown parameters.  Equation~(\ref{eq:tobesolved}) holds for each unit velocity bin if we redefine $N$ as the column density per unit velocity width.  Given the continuum covering factor $f_\mathrm{c}$, we can obtain optical depth profiles from absorption line profiles using equation~(\ref{eq:taudef}), and thus, the excitation temperature profile. 

Populations of rotational energy levels for molecules in interstellar space are determined by collisional and radiative processes.  In the absence of any luminous object nearby, the CMB dominates the radiative processes toward PKS1830--211 SW.  The relative contribution of collisional excitation is proportional to the density of colliding particles and can be assessed by comparison with the critical density.  The rotational transitions of molecules with large electric dipole moment have large critical densities.  For example, the critical density of HCN {\it J}=1--0 line is as high as $n_\mathrm{crit}({\rm H}_2)\!\sim\!{10^{6}}\,\mathrm{cm^{-3}}$ and even higher for the upper transitions.  The density and kinetic temperature of the PKS1830--211 SW absorber are estimated to be $n_\mathrm{H_2}\!\sim\!{2600}\,\mathrm{cm^{-3}}$ \citep{Henkel_2009} and $T_\mathrm{kin}\!\sim\!{80 \,\mathrm{K}}$ \citep{Henkel_2008}, respectively.  Further, the local radiation field surrounding the absorber is reported to be negligible \citep{Muller_2013}.  Thus, we can conclude that, in this SW absorber, the rotational level populations of highly polar molecules (HCN, HCO$^{+}$, CS etc.) are in equilibrium with the CMB radiation.  The situation assures that the approximate equation 
\begin{equation}\label{eq:tcmb=tex}
T_\mathrm{CMB} = T_\mathrm{ex}
\end{equation}
holds.  The previous measurements of $T_\mathrm{CMB}$ from absorption lines of highly polar molecules toward the PKS1830--211 SW assume this relationship \citep{Wiklind_1996,Menten_1999}.  The analyses in this study follow this as well.  

Another powerful approach to constraining the CMB temperature is non-LTE analysis with the RADEX code \citep{vanderTak_2007}. We note, however, that RADEX calculations under optically thin conditions exhibit anomalous behavior for higher-$J$ transitions, as detailed in Appendix \ref{appendix}. For this reason, we did not use RADEX outputs as the basis for systematic error estimates in this study.

\subsection{Uncertainty}\label{subsec:uncertainty}
If equation~(\ref{eq:taudef}) holds, the uncertainty in the measured $T_{\mathrm{ex}}$ arises from the indeterminacy of $f_\mathrm{c}$ and temporal variations in the absorption strength.

\subsubsection{Continuum Covering Factor} \label{subsubsec:fc}
An inappropriate assumption of $f_\mathrm{c}$ leads to the incalculability of optical depth, especially in optically thick cases.  The value of $f_\mathrm{c}$ in SW is evaluated from the bottoms of saturated absorption lines, which ranges from 0.91--0.97 \citep[][]{Muller_2014,Muller_2016,Muller_2017} with no widely accepted value yet.  Some studies assume $f_\mathrm{c}\!=\!1$ for simplicity \citep[e.g.,][]{Wiklind_1998, Muller_2013, Muller_2020, Tercero_2020}, and thus, the results are subject to no small errors. Moreover, $f_\mathrm{c}$ is known to vary with both time and frequency \citep[e.g.,][]{Muller_2014, Muller_2021, Muller_2023}.

We employed a Monte Carlo approach to rigorously assess these uncertainties.  We assumed that $f_\mathrm{c}$ follows a uniform distribution over the range 0.90--1.00, and that each observed data point $I_\mathrm{obs} \equiv I_\mathrm{c} - \Delta I$ follows a normal distribution characterized by the observed value as its mean and the associated uncertainty as its standard deviation. Random samples of $f_\mathrm{c}$ and $I_\mathrm{obs}$ were independently drawn from their respective distributions for each velocity bin and transition, and the optical depth was calculated using equation~(\ref{eq:taudef}).  This process was repeated 100,000 times to construct the probability distribution of the optical depth. Subsequently, the median and 1$\sigma$ confidence interval of the resulting distribution were adopted as the observed optical depth ($\tau_\mathrm{obs}$) and its corresponding uncertainty, respectively.  This method enables the joint propagation of uncertainties in $f_\mathrm{c}$, accounting for its temporal and frequency dependence, as well as statistical uncertainties in $I_\mathrm{obs}$, into the final estimate of $\tau_\mathrm{obs}$.

\subsubsection{Time Variability}\label{subsubsec:time_var}
Temporal variability in molecular absorption profiles toward PKS1830--211 has been reported in several studies \citep[e.g.,][]{Muller_2008, Muller_2014, Muller_2020, Muller_2023, Schulz_2015}. The ALMA data used in this study span multiple epochs between 2014 and 2019. 
Notably, the HCN {\it J}=4--3 and {\it J}=5--4 spectra were observed on July 28 and April 11, 2019, respectively---during a period when significant temporal variations were observed in the absorption profiles of molecules such as $\mathrm{H_2^{18}O}$, $\mathrm{CH_3OH}$, $\mathrm{N_2H^+}$, and $\mathrm{H_2CO}$ \citep{Muller_2020}. 
\citet{Muller_2020} reported a flare event of PKS1830--211 over this period, which is considered to be responsible for the observed variations. They further showed that, during the flare, the integrated optical depth of the blueshifted component at $\sim-5\,\mathrm{km\,s^{-1}}$ increased by a factor of two, while the redshifted component at $\sim+5\,\mathrm{km\,s^{-1}}$ remained nearly constant. Since no pre-flare observations of the HCN {\it J}=4--3 and {\it J}=5--4 profiles toward PKS1830--211 SW are available, we adopted this variability, observed in the optically thin molecular lines, as a proxy for potential flare-induced variability in these transitions and assessed its impact on the uncertainty of the excitation temperatures.

Based on the results of \citet{Muller_2020}, the integrated optical depth derived from our flare-epoch HCN {\it J}=4--3 and {\it J}=5--4 spectra could decrease by about a factor of two during quiescent periods. We therefore assigned a systematic downward uncertainty to the integrated optical depth in the velocity range $V_\mathrm{bar}\!=\!-10$ to $0\,\mathrm{km\,s^{-1}}$. Specifically, for each velocity bin $i$ with integrated optical depth $S_i$, we adopted a systematic uncertainty of $\sigma_{\mathrm{sys},i}=0.5S_i$. The total downward uncertainty was then given by $\sqrt{\sigma_{\mathrm{sta},i}^2 + \sigma_{\mathrm{sys}, i}^2}$, where $\sigma_{\mathrm{sta},i}$ is the statistical uncertainty. The upward uncertainty was taken to be purely statistical. Using the integrated optical depths with these combined uncertainties, we calculated the excitation temperatures and their uncertainties in each velocity bin (Sect.~\ref{sec:Tex}).

\subsection{Correction of Non-Uniformity}\label{subsec:tau_cor}
Equation~(\ref{eq:taudef}) assumes a cloud with uniform column density partially covering the background continuum source, whereas actual clouds have non-uniform column densities.  The contribution of high column density areas to $\Delta I$ tends to saturate, and therefore, the effect of non-uniformity yields a downward revision to $\tau_{\nu}$ and $N_{J}$.  This effect becomes more serious the higher the optical depth, which results in the overestimation of $T_{\rm ex}$.  

\begin{figure}[htbp]   \centering
  \includegraphics[width=\linewidth]{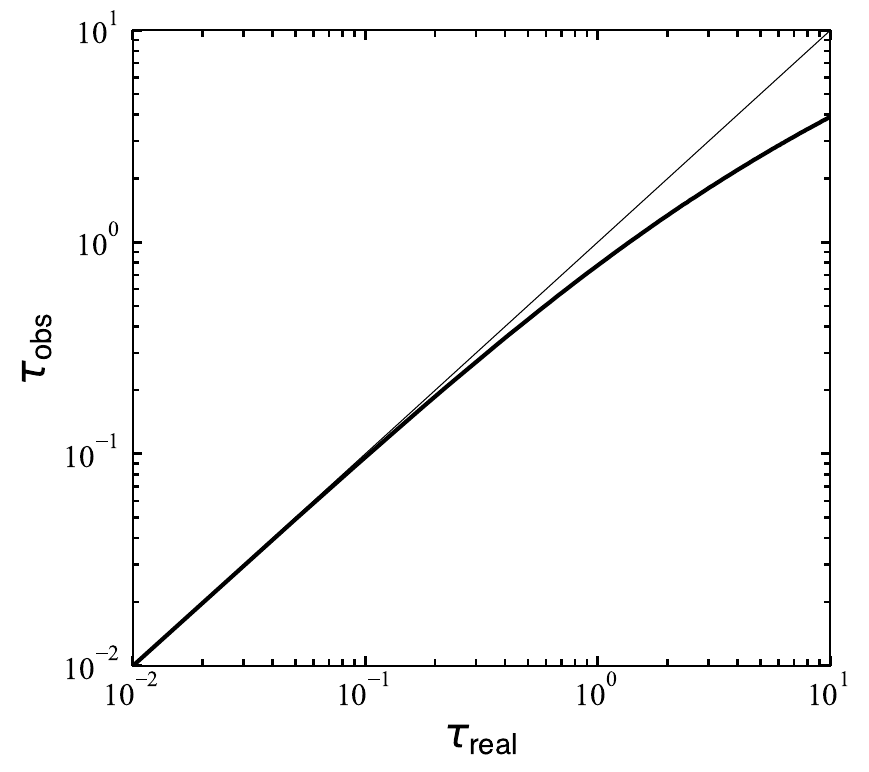}
  \caption{Calculated relationship between $\tau_{\rm obs}$ and $\tau_{\rm real}$, assuming the N-PDF determined in the Orion A giant molecular cloud.  The thin solid line denotes $\tau_{\rm obs}=\tau_{\rm real}$.}
  \label{fig:tau_cor}
\end{figure}

Assuming the lognormal probability distribution function for column density (N-PDF), we calculated the correction factor between the observed optical depth $\tau_{\rm obs}$ and averaged actual optical depth ($\tau_{\rm real}$).  The parameters of N-PDF determined in the Orion A giant molecular cloud \citep{Ma_2020} were employed for our calculation because its average column density ($N({\rm H}_2)\!\simeq\! 1.5\!\times\! 10^{22}\,{\rm cm}^{-2}$) is comparable to that of the $z\!=\!0.89$ absorber.  Figure \ref{fig:tau_cor} shows the calculated relationship between $\tau_{\rm obs}$ and $\tau_{\rm real}$, which is used for our $T_{\rm ex}$ calculations.  This correction of non-uniformity was applied to all data points.  

\section{Results and Discussion} \label{sec:resdis}
\subsection{Optical Depth Profiles} \label{sec:opa}
\begin{figure}[htbp]   \centering
  \includegraphics[width=\linewidth]{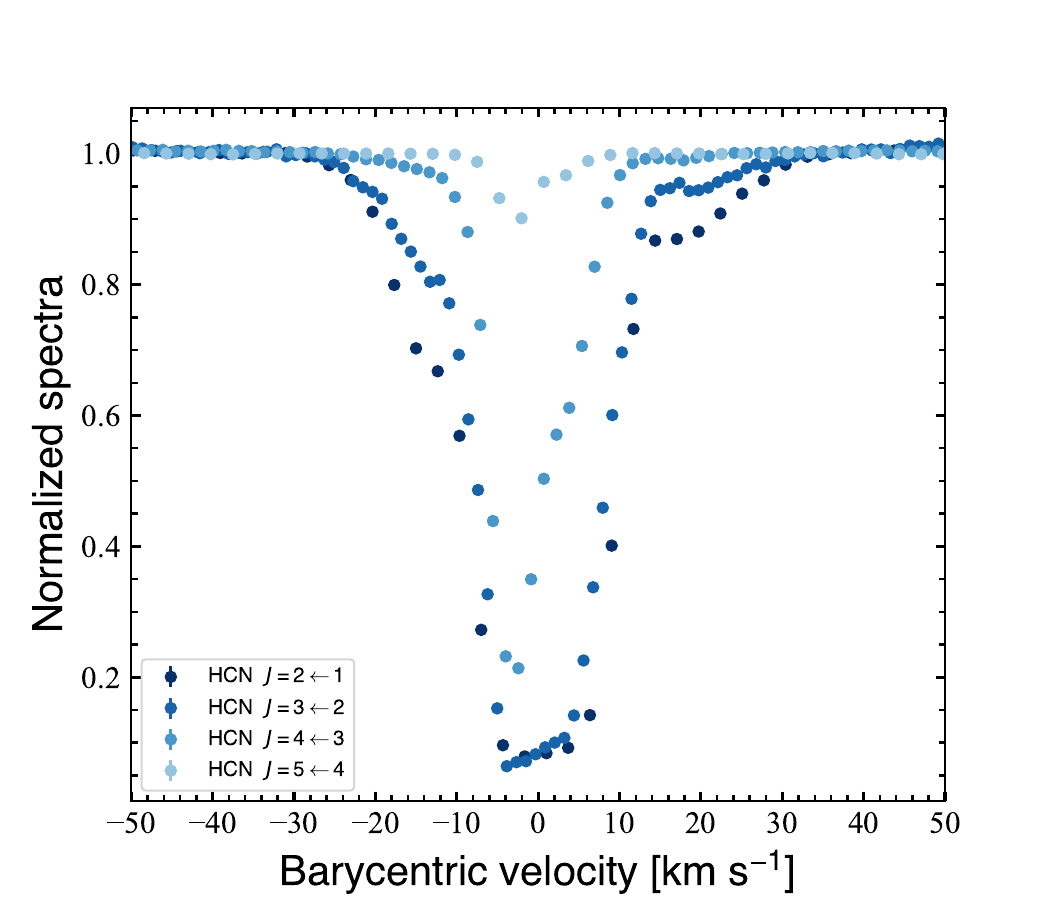}
  \caption{Continuum normalized absorption spectra for all HCN transitions toward PKS1830--211 SW. The velocities are offsets from $V_\mathrm{sys}$ in the barycentric frame.}
  \label{fig:absprofile}
\end{figure}

Figure \ref{fig:absprofile} shows the normalized absorption spectra of HCN {\it J}=2--1, {\it J}=3--2, {\it J}=4--3 and {\it J}=5--4 lines toward PKS1830--211 SW. The top panel of Figure \ref{fig:Texprofile} presents the bias-corrected optical depth spectra, accounting for the indeterminacy of $f_{\rm c}$.
We clearly see at least four components centered at $V_{\rm bar}\!=\! -13$, $-3$, $+3$, and $+17\,\mathrm{km\,s^{-1}}$.  The {\it J}=2--1 and {\it J}=3--2 absorption profiles of the central two components are fully saturated at each velocity center, which makes calculated optical depths highly uncertain.  The values of these saturated bottoms provide the continuum covering factor of $f_{\rm c}\!=\!0.91\mbox{--}0.92$ for the central two components.  

Previous measurements of $T_\mathrm{CMB}$ based on molecular absorption lines used velocity-integrated absorption intensities of less opaque species \citep[e.g.,][]{Muller_2013}.  This method could avoid uncertainty in optical depths because of the saturation of absorption, but it is subject to relatively large statistical errors.  In this study, we took a different approach, employing an opaque species, evaluating optical depths at each velocity channel, and taking a weighted mean of the calculated $T_\mathrm{ex}$.  

\subsection{Excitation Temperature Profile} \label{sec:Tex}
Absorption spectra were resampled onto the velocity grid of the {\it J}=5--4 data ($\Delta V_\mathrm{bar}=2.728\,\mathrm{km\, s^{-1}}$), which has the lowest velocity resolution.  The excitation temperature $T_\mathrm{ex}$ and column density $N$ were determined by the weighted least-squares-fitting of equation~(\ref{eq:tobesolved}) to the calculated optical depths.  The initial values of $T_\mathrm{ex}$ and $N$ were set to the standard Big Bang model prediction of $T_\mathrm{CMB}(z = 0.89) = 5.14\,\mathrm{K}$ and $N = 3.041 \times 10^{14}\,\mathrm{cm^{-2}}$, as reported by \citet{Muller_2011}.  Only optical depths lower than unity---where photon trapping and collisional excitation are expected to be negligible and equation (\ref{eq:tcmb=tex}) remains valid---were used in the fitting.

\begin{figure}[htbp] 
  \centering
  \includegraphics[width=\linewidth]{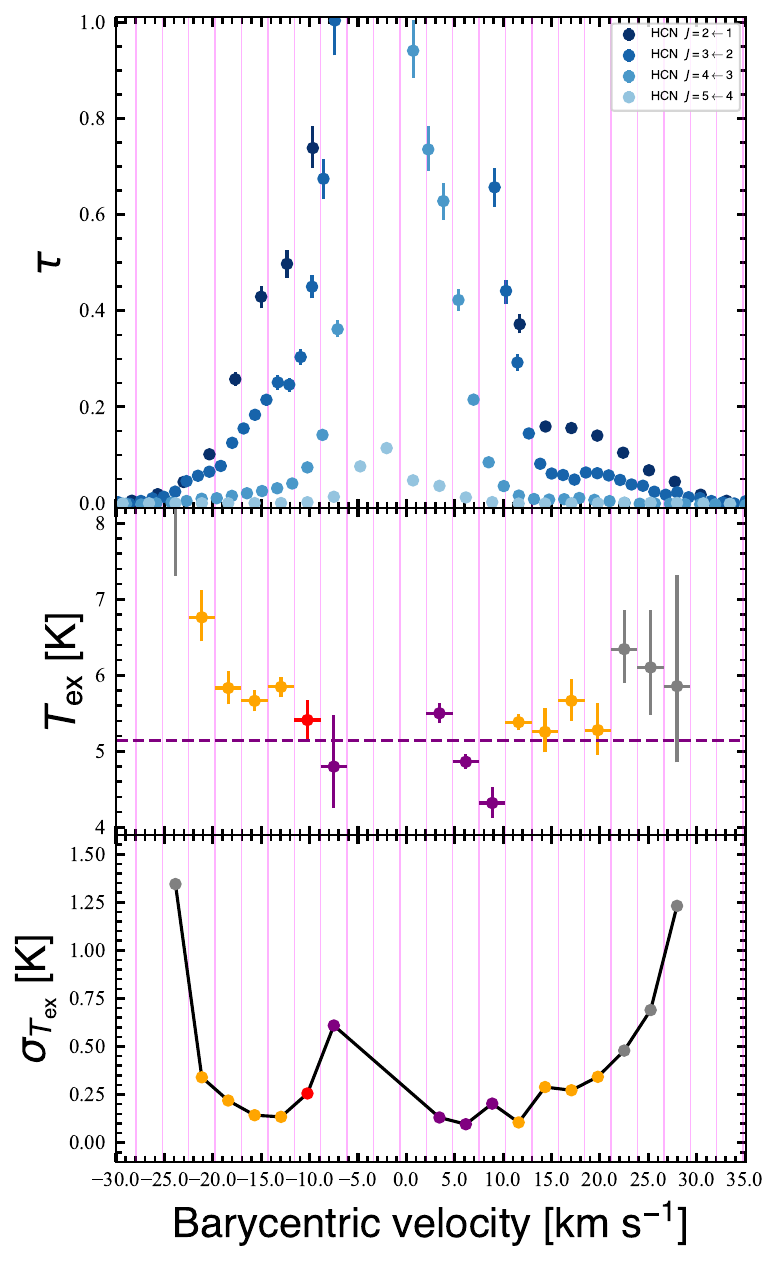}
  \caption{Profiles after optical depth correction. (Top) The optical depth profiles of all HCN transitions toward PKS1830--211 SW.  The velocity range is limited to the region with significant absorption detected in two or more transitions.  The vertical thin lines indicate the boundaries of velocity bins.  (Middle) The excitation temperature profile calculated from optical depths lower than unity.  The purple dashed line denotes $T_{\rm ex}\!=\!5.14\,\mathrm{K}$, which is the CMB temperature at $z\!=\!0.89$ expected from the standard Big Bang model.  (Bottom) The uncertainty in $T_\mathrm{ex}$ was defined as the average of the upper and lower deviations from the median of the Monte Carlo-derived distribution. In the middle and bottom panels, colors represent the $J$ transition sets used to derive $T_{\rm{ex}}$ and $N$: gray for (2--1, 3--2), orange for (2--1, 3--2, 4--3), red for (2--1, 3--2, 4--3, 5--4) , and purple for (4--3, 5--4).}
  \label{fig:Texprofile}
\end{figure}

The middle and lower panel in Figure \ref{fig:Texprofile} present the excitation temperature and uncertainty profiles, respectively,  accounting for the uncertainty in $f_{\rm c}$ and temporal variations in absorption depth.  The combination of transitions used in the $T_{\rm ex}$ calculation is indicated by color.  The uncertainties in $T_{\rm{ex}}$ are asymmetric and correspond to the 68\% confidence intervals derived from the Monte Carlo sampling.
The standard error ($\sigma_{T_{\rm{ex}}}$) was estimated as the average of the upper and lower bounds.

Based on the excitation temperature profile shown in Figure \ref{fig:Texprofile}, we focus on three velocity ranges: $V_\mathrm{bar} = -13\,\mathrm{km\,s^{-1}}$, $-3$ to $+3\,\mathrm{km\,s^{-1}}$, and $+17\,\mathrm{km\,s^{-1}}$. In the $-3$ to $+3\,\mathrm{km\,s^{-1}}$ component, the $J$=2--1 and $J$=3--2 absorption lines are saturated, allowing the covering factor to be constrained to $f_{\mathrm{c}}=0.9$--$1.0$. By contrast, the $-13\,\mathrm{km\,s^{-1}}$ and $+17\,\mathrm{km\,s^{-1}}$ components show no saturation, leaving their covering factors unconstrained and the derived excitation temperatures uncertain. Consequently, we regard the temperatures obtained for the $-3$ to $+3\,\mathrm{km\,s^{-1}}$ component ($-12 \leq V_{\mathrm{bar}} \leq +13\,\mathrm{km\,s^{-1}}$ ) as the most reliable estimate of the CMB temperature.
For the resulting $T_{\rm ex}$, we calculated the mean value by
\begin{equation}\label{meanTex}
    \left<T_{\rm ex} \right> \equiv \sum{\frac{T_{\rm ex}}{\sigma_{T_\mathrm{ex}}^2}}/\sum{\frac{1}{\sigma_{T_\mathrm{ex}}^2}} .
\end{equation}
The final value obtained is $\left<T_{\rm ex} \right>\!=\!5.13\pm0.06\,\mathrm{K}$, which may be a good approximation of $T_{\rm CMB}$ at the $z\!=\!0.89$ absorber toward PKS1830--211 SW.

\subsection{Redshift Dependence of $T_\mathrm{CMB}$} \label{sec:cosmo}
Under the assumption of the CMB-dominated excitation condition for HCN rotational energy levels, we obtained $T_\mathrm{CMB}(z\!=\!0.89)\!=\!5.13\pm0.06 \,\mathrm{K}$.  This is in good agreement with the value ($5.14 \,\mathrm{K}$) expected from the standard Big Bang model within the uncertainty.  The uncertainty is reduced by $\sim\!40$\% relative to the previous best estimate of \citet{Muller_2013}.  Our new precise measurement of $T_\mathrm{CMB}$ at $z\!=\!0.89$ yields constraints on deviations from the standard model, consistent with ealier work.

\begin{figure}[htbp]
  \centering
  \includegraphics[width=\linewidth]{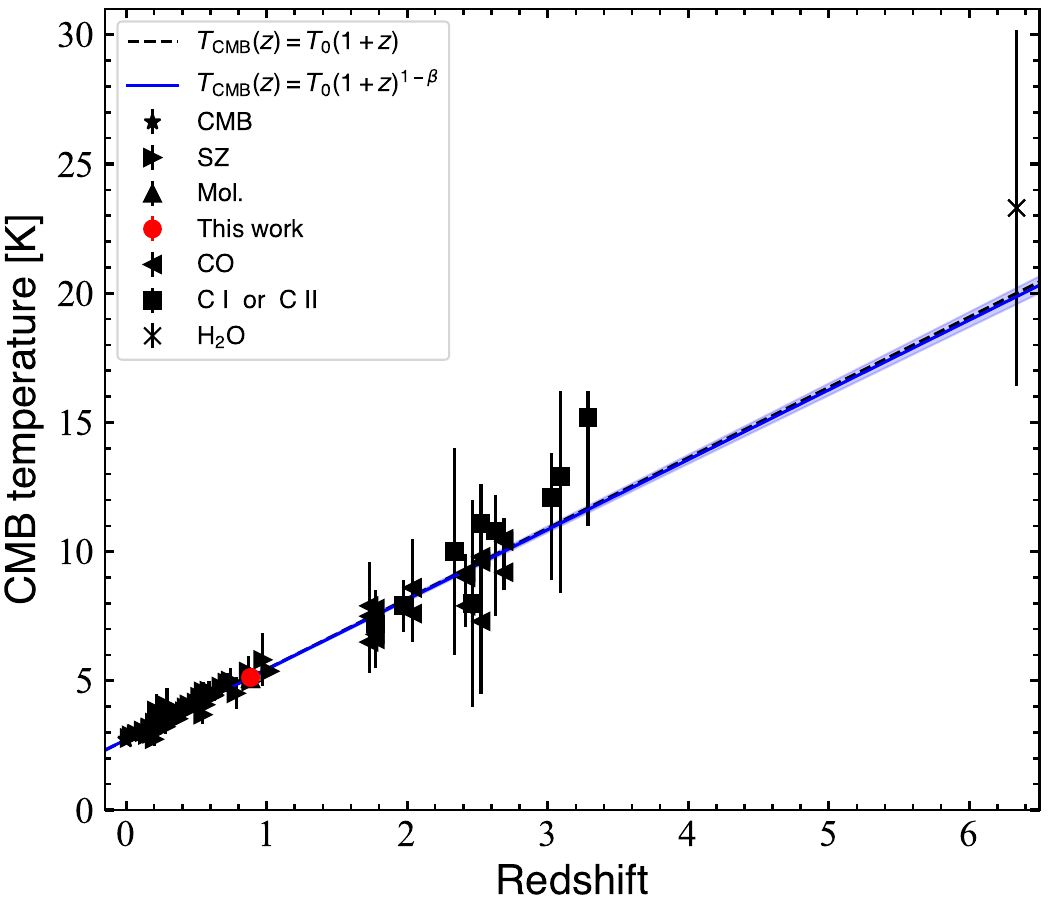}
   \caption{Plot of the measured CMB temperature ($T_\mathrm{CMB}$) versus redshift ($z$).  The values of previous measurements were taken from existing literature \citep[][and references therein]{Battistelli_2002, Luzzi_2009, Muller_2013, Klimenko_2020, Riechers_2022}.  A star indicates the measurement at $z\!=\!0$ \citep{Fixsen_2009}.  Each plot symbol represents a measurement method at $z>0$.  The black dashed line denotes the relationship expected from the standard model [eq.(\ref{eq:tcmb})].  The blue curve shows the best-fit result by the eq.(\ref{eq:btcmb}), $\beta=(3.2^{+7.5}_{-8.4})\times10^{-3}$, with the 1$\sigma$ uncertainty area (blue shadow).}
   \label{fig:z-Tcmb}
\end{figure}

Figure \ref{fig:z-Tcmb} shows a plot of the measured CMB temperature versus the cosmological redshift.  The results from about 80 measurements of $T_\mathrm{CMB}(z)$ \citep[][and references therein]{Battistelli_2002, Luzzi_2009, Muller_2013, Klimenko_2020, Riechers_2022} were plotted with that from this study.  The $T_\mathrm{CMB}\mbox{-}z$ plot was fitted by the model relationship \citet{Lima_2000}, 
\begin{equation}\label{eq:btcmb}
T_\mathrm{CMB}(z)=T_0\,(1+z)^{1-\beta} , 
\end{equation}
which yields ${\beta=(3.2^{+7.5}_{-8.4})\times10^{-3}}$. This result indicates that the $T_\mathrm{CMB}(z)$ measurements obtained so far are still compatible with the standard Big Bang model. 
Our measurement places a robust constraint on deviations from the model, consistent with and complementary to the recent value of $\beta=(3.4^{+8.1}_{-7.3})\times10^{-3}$ reported by \citet{Riechers_2022}.

\section{Summary and Future Prospects}
In this study, the millimeter-wave data toward PKS1830--211 observed with the ALMA were analyzed to calculate the most reliable value for the CMB temperature at $z\!=\!0.89$.  We obtained the absorption spectra of four lowest-{\it J} transitions of HCN and calculated the profiles of $T_{\rm ex}$ and its uncertainty.  The effect of uncertainty in the continuum covering factor was evaluated using a Monte Carlo approach over the expected range. We further considered temporal variations in the integrated optical depths during the flare, from which we estimated the excitation temperatures and their associated uncertainties.  The bias in the observed optical depth attributed to the non-uniformity of the absorbing cloud was corrected assuming a lognormal probability distribution for the column density.  Taking the weighted mean of calculated $T_{\rm ex}$, we obtained $T_\mathrm{CMB}(z\!=\!0.89)\!=\!5.13\pm0.06 \,\mathrm{K}$, in good agreement with the value predicted by the standard Big Bang model, thus yielding constraints on non-standard models that is consistent with previous determinations.

Direct quantification of uncertainties associated with temporal variability and covering factors in the multiple subcomponents of the HCN spectra remains a key challenge for future studies. Addressing this will require high-resolution spectral monitoring capable of resolving these subcomponents.

We also note that RADEX outputs under optically thin conditions show unphysical trends (Appendix \ref{appendix}), and were therefore not adopted in this work.

More precise measurements of $T_\mathrm{CMB}$ at redshifts higher than unity are essential for a more rigorous examination of the standard model.  Millimeter-wave observations of the [\Cone] fine-structure lines with the ALMA toward quasars with $z\!=\!2\mbox{--}3$ absorbers could be the next targets.  New generation telescopes such as the square kilometer array and the next generation very large array, as well as the upgraded ALMA, are expected to significantly increase the number and accuracy of $T_\mathrm{CMB}$ measurements at $z\!>\!3$, enabling us to search for any slight deviation from the standard model.

%% Please use the acknowledgment and contribution environments. This will be anonomyized when the "anonymous" style option is used. 
\begin{acknowledgments}
  This paper makes use of the following ALMA data: ADS/JAO.ALMA\#2013.1.01099.S, ADS/JAO.ALMA\#2017.1.01119.S, ADS/JAO.ALMA \#2018.1.00051.S, and ADS/JAO.ALMA\#2018.1.00692.S. ALMA is a partnership of ESO (representing its member states), NSF (USA) and NINS (Japan), together with NRC (Canada), MOST and ASIAA (Taiwan), and KASI (Republic of Korea), in cooperation with the Republic of Chile.  The Joint ALMA Observatory is operated by ESO, AUI/NRAO and NAOJ.  T.K. acknowledges support from the Nordic ALMA Regional Centre (ARC) node based at Onsala Space Observatory.  The Nordic ARC node is funded through Swedish Research Council grant No 2019-00208.  T.O. acknowledges the support from JSPS Grant-in-Aid for Scientific Research (A) No. 20H00178.  We would like to thank Editage (\url{www.editage.jp}) for English language editing.
\end{acknowledgments}

%% To help institutions obtain information on the effectiveness of their  telescopes the AAS Journals has created a group of keywords for telescope 
%% Following the acknowledgments section, use the following syntax and the \facility{} or \facilities{} macros to list the keywords of facilities used in the research for the paper.  
\facility{ALMA.}

%% Similar to \facility{}, there is the optional \software command to allow authors a place to specify which programs were used during the creation of the manuscript. Authors should list each code and include either a citation or url to the code inside ()s when available.
\software{CASA \citep{CASATeam_2022}, Matplotlib \citep{Hunter_2007}, NumPy (\citealp{vanderWalt_2011, Harris_2020}), SciPy \citep{Virtanen_2020}, and UVMULTIFIT \citep{MartiVidal_2014}.}

\appendix

\section{RADEX Output Behavior in the Optically Thin Limit} \label{appendix}

As noted in the main text (Sect. \ref{subsec:calculation}), the behavior of RADEX in the optically thin limit requires careful consideration. Detailed results are presented here for completeness.  We examined RADEX outputs under extremely optically thin conditions, using $\mathrm{H_2}$ as the collision partner. Collisional rate coefficients were taken from the Leiden Atomic and Molecular Database \citep[LAMDA;][]{Schoier_2005}. The background radiation temperature was set to $T_{\mathrm{CMB}} = 5.14$ K, the kinetic temperature to $T_{\mathrm{kin}} = 80$ K, and the H$_2$ density was varied over $n_{\mathrm{H_2}} = 10^2$--$10^4$ cm$^{-3}$. The column density per velocity width was set to $N/\Delta V = 10^{11}$--$10^{13}$ cm$^{-2}$ (km s$^{-1}$)$^{-1}$. HCN rotational transitions from $J$=1--0 through $J$=8--7 were considered.

\begin{figure}[htbp]
  \centering
  \includegraphics[width=\linewidth]{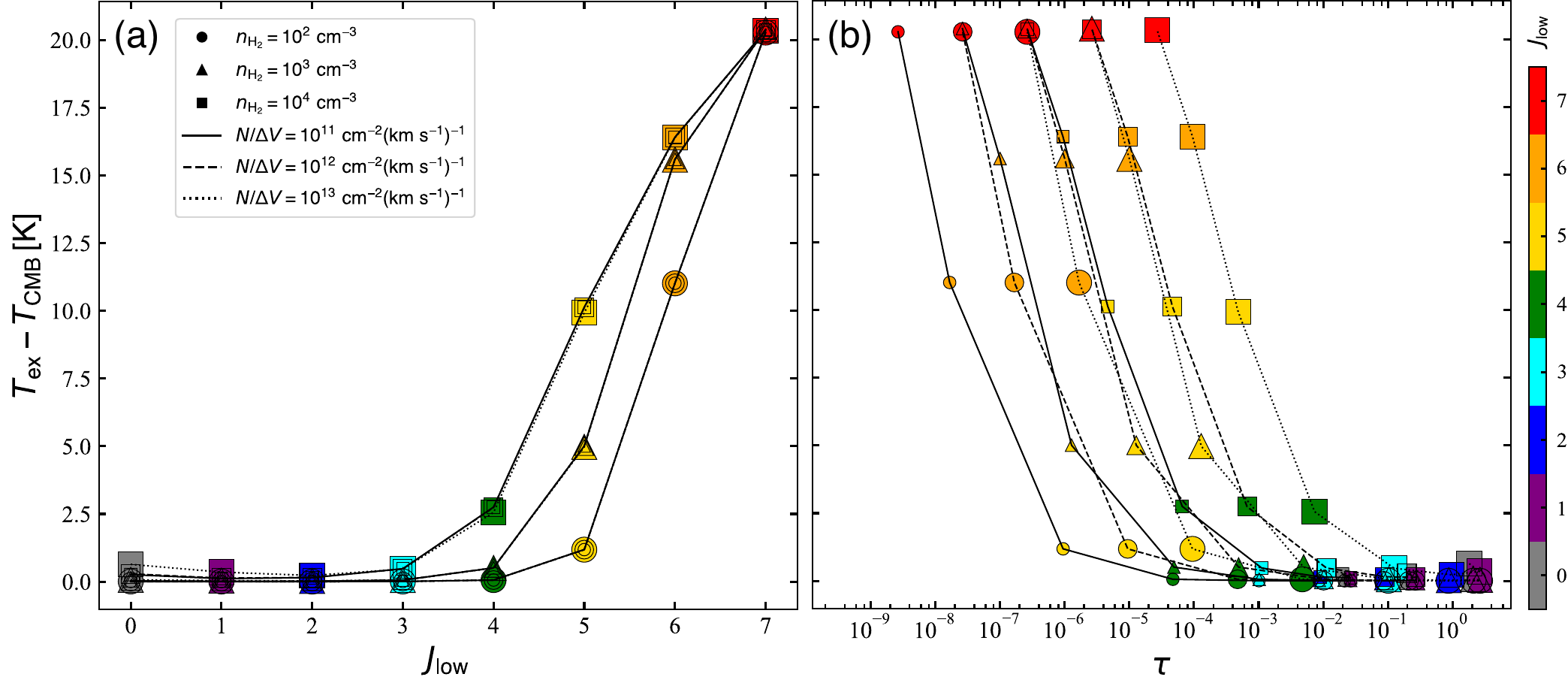}
  \caption{(a) Deviation of the excitation temperature ($T_{\mathrm{ex}}$) from $T_{\mathrm{CMB}}$ as a function of rotational quantum number $J$ for HCN transitions from $J$=1--0 to $J$=8--7. Solid, dashed, and dotted lines with circular, triangular, and square markers show RADEX outputs for the optically thin models described in the text. Marker shapes denote different $\mathrm{H_2}$ number densities, while marker sizes scale with $N/\Delta V$, such that larger markers correspond to higher $N/\Delta V$ values. The systematic but unphysical increase of $T_{\mathrm{ex}}$ with $J$, starting at $J \geq 4$, is clearly evident. (b) The deviation as a function of optical depth ($\tau$) for the same model parameters as in panel (a). Even at $\tau \ll 10^{-3}$, $T_{\mathrm{ex}}$ departs significantly from $T_{\mathrm{CMB}}$ at higher-$J$ transitions, contrary to expectations from basic radiative transfer principles.}
  \label{fig:Tex}
\end{figure}

Basic radiative transfer predicts that, in the limit of very low optical depth ($\tau \ll 1$) and negligible collisional excitation, the excitation temperature $T_{\mathrm{ex}}$ should converge to $T_{\mathrm{CMB}}$, independent of $J$.
The RADEX results, however, display clear anomalies. For intermediate- and high-$J$ transitions ($J \geq 4$), $T_{\mathrm{ex}}$ increases systematically with $J$ (Figure \ref{fig:Tex}(a)), even when $\tau \ll 10^{-3}$ (Figure \ref{fig:Tex}(b)). While $J$=1--0 and 2--1 yield $T_{\mathrm{ex}} \approx T_{\mathrm{CMB}}$ as expected, the $J$=4--3 and 5--4 transitions show departures of 0.1--several K above $T_{\mathrm{CMB}}$, with larger deviations for $J \geq 6$. These trends are inconsistent with the negligible role of collisional excitation in the adopted density range and contradict the expected $J$-independent convergence to $T_{\mathrm{CMB}}$.

This behavior is inconsistent with basic radiative transfer in the optically thin limit, particularly at higher $J$, and casts significant doubt on the reliability of RADEX predictions under such conditions. Therefore, we did not adopt non-LTE analysis based on RADEX calculations in the present study.

%% For this sample we use BibTeX plus aasjournalv7.bst to generate the

% \bibliography{sample701}{}
% \bibliographystyle{aasjournalv7}

\end{document}